\documentstyle[seceq]{ptptex}

\title{
Can We See a Rotating Gravitational Lens? 
}

\author{
Hideki {\sc Asada}$^{1,2,}$\footnote{E-mail: 
asada@phys.hirosaki-u.ac.jp} 
and Masumi {\sc Kasai}$^{1,}$\footnote{E-mail: 
kasai@phys.hirosaki-u.ac.jp} 
}

\inst{%
$^1$ Faculty of Science and Technology, Hirosaki University, 
Hirosaki 036-8561, Japan\\
$^2$ Max-Planck-Institut f\"ur Gravitationsphysik 
(Albert-Einstein-Institut), Am M\"uhlenberg 5, D-14476 Golm, Germany
}

\abst{
It is known that the rotation of a gravitational lens affects 
properties of images. 
We consider an inverse problem: If the lens is dark, can we infer 
its rotation from the observed images? 
We find that, up to the first order in the gravitational constant $G$, 
a rotating lens is not distinguishable from a non-rotating one. 
The quadrupole contribution to the deflection angle is also calculated 
for a general, extended lens object. 
}

\begin{document}

\maketitle

\section{Introduction}

The gravitational lens is an important phenomenon in astrophysics 
for probing mass distributions. \cite{SEF}  
Most studies of gravitational lensing have assumed that 
the lens object is non-rotating. 
However, celestial objects such as stars and galaxies are 
usually rotating. 
Hence, it is important to consider the prospect of  
probing the rotation through the gravitational lensing. 
In particular, recent rapid advances in observations 
enable us to see the light passing in the vicinity of neutron stars, 
black holes, and so on. 
For this reason, we may obtain a great deal of useful information 
regarding lens objects. 

As for the sun, Epstein and Shapiro estimated the deflection angle 
due to its spin in the context of the parameterized 
post-Newtonian (PPN) approximation of gravitational theories. \cite{ES} 
For more general lens models, observable effects due to the rotation 
have been discussed by several authors. \cite{Ibanez,Sari,KK} 
On the other hand, in the study of the lensing of gravitational waves, 
Baraldo et al. have found that an interference pattern is shifted 
translationally in proportional to the Kerr parameter without 
any change in the shape of the pattern. \cite{BHN} 
We consider an inverse problem: If the lens is dark, can we infer 
its rotation from the observed images?  

The conventional derivation of the lens equation is based on 
the non-rotating assumption. 
Hence, in order to clarify the effect due to rotation, we first 
summarize the general formalism of the gravitational lens in Section 2. 
We also obtain the lens equation for a rotating extended object 
that produces a weak gravitational field along the line of sight. 
In Section 3, we use the multipole expansion in order to take 
account of the effect due to the extended nature of lens. 
In particular, the quadrupole contribution, which does not appear 
in the linearized Kerr metric, is also studied. 
Section 4 is devoted to a summary and conclusion. 

\section{General formulation of the gravitational lens}

First, we summarize the notation and give the equations for gravitational
lensing. We basically employ the notation of Ref. \cite{SEF}, but the 
signature is $(-,+,+,+)$.

\subsection{The $3+1$ splitting of stationary spacetime}

The metric of stationary spacetime can be written in the form 
\begin{equation}
  \label{eq:llform}
  ds^2 = g_{\mu\nu} dx^{\mu}\,dx^{\nu} = 
   - h \left( cdt- w_i dx^i \right)^2 + h^{-1} \gamma_{ij} dx^i\,dx^j, 
\end{equation}
where 
\begin{equation}
  \label{eq:components}
  h \equiv - g_{00}, \quad w_i \equiv - \frac{g_{0i}}{g_{00}},  
\end{equation}
and 
\begin{equation}
  \label{eq:gammaij}
  \gamma_{ij} dx^i\,dx^j \equiv -g_{00}
  \left(g_{ij} - \frac{g_{0i}\, g_{0j}}{g_{00}}\right)dx^i\,dx^j
  \equiv d\ell^2. 
\end{equation}
This is essentially the same as the Landau-Lifshitz $3+1$
decomposition of stationary spacetime. \cite{LL}
The only difference here is in the definition of the spatial metric. 
In the Landau-Lifshitz case, this is 
\begin{equation}
  \label{eq:llgamma}
  \tilde{\gamma}_{ij} \equiv \left(g_{ij} - \frac{g_{0i}\,
  g_{0j}}{g_{00}}\right)
  = h^{-1} \gamma_{ij} . 
\end{equation}
We will hereafter use a conformally rescaled 
$\gamma_{ij}$ for the following reason: 
The spatial distance $d\ell$ defined by Eq.~(\ref{eq:gammaij}) behaves 
as the affine parameter of the null geodesics in this spacetime, 
as shown in the Appendix.  
Therefore, it is convenient to use $\gamma_{ij}$ when investigating
light ray propagation. 
The conformal factor $h$ corresponds to the gravitational redshift
factor.

\subsection{Fermat's principle in stationary spacetime}

For a future-directed light ray, the null condition $ds^2 = 0$ gives
\begin{equation}
  c\, dt = \frac{1}{h} \sqrt{\gamma_{ij}dx^i dx^j}
        + w_i \, dx^i . 
\label{eq:time}
\end{equation}
Since the spacetime is stationary, $h, \gamma_{ij}$ and $w_i$ are
functions of the spatial coordinates $x_i$ only. 
Then, the arrival time of a light ray is given by the integral of 
Eq.~(\ref{eq:time}) from the source to the observer (denoted by the 
subscripts $S$ and $O$, respectively):  
\begin{equation}
  t \equiv \int_{t_S}^{t_{O}} dt = \frac{1}{c} \int_{S}^{O}
  \left(\frac{1}{h}\sqrt{\gamma_{ij}e^i e^j}
        + w_i \,e^i\right) d\ell.  
\end{equation}
Here
$e^i = {dx^i}/{d\ell}$ is the unit tangent vector, 
which represents the direction of the light ray. 
Hereafter, lowering and raising the indices of the
spatial vectors are accomplished by $\gamma_{ij}$ and 
its inverse $\gamma^{ij}$. 
We note here that the quantity 
\begin{equation}
  n \equiv \frac{1}{h}\sqrt{\gamma_{ij} e^i e^j}
        + w_i \,e^i
\end{equation}
acts as an effective refraction index. 

Now Fermat's principle \cite{Perlick} states that $\delta t = 0$. 
{}From the Euler-Lagrange
equation, we obtain the equation for the light rays as follows:
\begin{equation}\label{eq:fulldedl}
  \frac{de^i}{d\ell} = - \left(\gamma^{ij} - e^i e^j\right) \partial_j 
  \ln h - \gamma^{il}\left(\gamma_{lj,k} - \frac{1}{2}
  \gamma_{jk,l}\right) e^j e^k + h \gamma^{ij}
  \left(w_{k,j}-w_{j,k}\right) e^k. 
\end{equation}
This ``equation  of motion'' for light rays is valid in any
stationary spacetime.

\subsection{Post-Minkowskian metric}

To this point, the treatment is fully exact. Hereafter, we use an 
approximate metric to represent the mildly inhomogeneous spacetime in
which the gravitational lensing events take place. We assume that the
perturbed spacetime is approximately described by the following
post-Minkowskian metric:
\begin{equation}
  \label{eq:pM}
  ds^2 = -\left( 1 + \frac{2\phi}{c^2}\right) c^2 dt^2
  + 2 c dt \frac{\psi_i \,dx^i}{c^3}
  + \left( 1 - \frac{2\phi}{c^2}\right)\delta_{ij} dx^i dx^j. 
\end{equation}
The energy-momentum tensor for a slowly moving, perfect fluid source 
 is
\begin{equation}
  \label{eq:tmunu}
  T^{00} \simeq \rho c^2, \quad T^{0i} \simeq \rho c v^i, \quad
  T^{ij} \simeq \rho v^i v^j + p \delta^{ij}. 
\end{equation}
In the near zone of a slowly moving, extended body, the metric is 
obtained up to the first-order in the gravitational constant $G$ from 
the integrals 
\begin{eqnarray}\label{eq:phi}
  \phi &=& -G \int \frac{\rho(\mbox{\boldmath $r$}^{\prime})}
{|\mbox{\boldmath $r$}-\mbox{\boldmath $r$}^{\prime}|}
  \,d^3 x^{\prime} , \\
  \psi_i &=& -4G \int \frac{\rho v_i (\mbox{\boldmath $r$}^{\prime})}
  {|\mbox{\boldmath $r$}-\mbox{\boldmath $r$}^{\prime}|} \,d^3
  x^{\prime}. 
\label{eq:psi}
\end{eqnarray}
Then, in the usual vector notation, the propagation equation 
(\ref{eq:fulldedl}) becomes \cite{YO,Durrer} 
\begin{equation}\label{eq:dedl}
  \frac{d\mbox{\boldmath $e$}}{d\ell} = - \frac{2}{c^2}
  \left(\nabla - \mbox{\boldmath $e$}\left(\mbox{\boldmath $e$}
  \cdot\nabla\right)\right)\phi + \frac{1}{c^3}\mbox{\boldmath $e$}
  \times(\nabla\times\mbox{\boldmath $\psi$}). 
\end{equation}

\subsection{The deflection angle}

The deflection angle $\mbox{\boldmath $\alpha$}$ is defined 
as the difference in the ray directions at the source and the observer, 
$\mbox{\boldmath $\alpha$} \equiv \mbox{\boldmath $e$}_{S} - 
\mbox{\boldmath $e$}_{O}$. Using Eq.~(\ref{eq:dedl}), we obtain
\begin{equation}\label{eq:alpha}
  \mbox{\boldmath $\alpha$} = - \int_S^O d\mbox{\boldmath $e$} = 
\int_S^O \left(
    \frac{2}{c^2} \left(\nabla -
      \mbox{\boldmath $e$}\left(\mbox{\boldmath $e$}\cdot\nabla\right)
    \right)\phi - \frac{1}{c^3}\mbox{\boldmath $e$}\times(\nabla\times
    \mbox{\boldmath $\psi$}) \right)
    d\ell. 
\end{equation}

\subsection{The lens equation}

The lens equation relates the image position  $\mbox{\boldmath $\xi$}$ 
to the source position $\mbox{\boldmath $\eta$}$ as 
\begin{equation}
  \mbox{\boldmath $\eta$} = \frac{D_{OS}}{D_{OL}}\,\mbox{\boldmath $\xi$}
    - D_{LS}\,\mbox{\boldmath $\alpha$}(\mbox{\boldmath $\xi$}),   
\end{equation}
where $D_{OS}$ is the distance from the observer to the source,
$D_{OL}$ is that from the observer to the lens, and $D_{LS}$ is 
that from the lens to the source. The vectors $\mbox{\boldmath $\xi$}$,
$\mbox{\boldmath $\eta$}$ and $\mbox{\boldmath $\alpha$}$ are 
2-dimensional vector in the sense that they are orthogonal to the ray 
direction $\mbox{\boldmath $e$}$ within our approximation. 
Alternatively, the lens equation is often written
in terms of the angular position vectors as 
\begin{equation}
  \mbox{\boldmath $\beta$} = \mbox{\boldmath $\theta$} - \frac{D_{LS}}{D_{OS}}
  \mbox{\boldmath $\alpha$}(D_{OL}\mbox{\boldmath $\theta$}), 
\end{equation}
where $\mbox{\boldmath $\beta$}=\mbox{\boldmath $\eta$}/D_{OS}$ 
is the unlensed position angle of the source and 
$\mbox{\boldmath $\theta$} = \mbox{\boldmath $\xi$}/D_{OL}$ is the
angular position of the image. In a cosmological situation, 
the unlensed position $\mbox{\boldmath $\beta$}$ 
(hence $\mbox{\boldmath $\eta$}$) is not an observable, 
because we cannot remove the lens from the observed position.

\section{Multipole expansion analysis}

The zeroth-order solution of the propagation equation 
(\ref{eq:dedl}) is $\mbox{\boldmath $e$} \equiv
\bar{\mbox{\boldmath $e$}}=\mbox{const}$. Therefore, 
within our approximation, the integration in Eq.~(\ref{eq:alpha}) 
is taken over the unperturbed ray,
$\mbox{\boldmath $x$}(\ell) = \mbox{\boldmath $\xi$} + 
\ell \bar{\mbox{\boldmath $e$}}$.  
Then Eq.~(\ref{eq:alpha}) can also be written as
\begin{equation}\label{eq:alpha2}
  \mbox{\boldmath $\alpha$} = \frac{2}{c^2}\int_S^O 
   \nabla\left(\phi - \frac{1}{2c}\bar{\mbox{\boldmath $e$}}\cdot
   \mbox{\boldmath $\psi$}\right)d\ell
  - \int_S^O (\bar{\mbox{\boldmath $e$}}\cdot\nabla) \left(
    \frac{2}{c^2}\bar{\mbox{\boldmath $e$}}\phi - \frac{1}{c^3} 
    \mbox{\boldmath $\psi$}\right)d\ell .
\end{equation}
The second part of the right-hand side of Eq.~(\ref{eq:alpha2}) is
simply written in terms of the boundary values. 
Its contribution is negligible in asymptotically flat regions. 
Consequently, the frame-dragging effect of a rotating lens can be 
expressed as the change of the Newton potential term in the following
way: 
\begin{equation}
  \phi \rightarrow \tilde{\phi} = \phi
    - \frac{1}{2c}\bar{\mbox{\boldmath $e$}}\cdot\mbox{\boldmath $\psi$}. 
\end{equation}
{}From Eqs.~({\ref{eq:phi}}) and ({\ref{eq:psi}}), we also have
\begin{equation}
  \label{eq:tildephi}
  \tilde{\phi}= -G \int
  \frac{\tilde{\rho}(\mbox{\boldmath $r$}^{\prime})}{|\mbox{\boldmath $r$}
  -\mbox{\boldmath $r$}^{\prime}|}
  \,d^3 x^{\prime}, 
\end{equation}
where 
\begin{equation}
  \tilde{\rho} \equiv \rho -
  \frac{2\rho \mbox{\boldmath $v$}\cdot\bar{\mbox{\boldmath $e$}}}{c}
\end{equation}
is the ``effective'' density for a rotating lens. Therefore, 
a rotating gravitational lens with density $\rho$ is equivalent to a
non-rotating one with effective density $\tilde{\rho}$. 

Under the assumption $|\mbox{\boldmath $r$}| > 
|\mbox{\boldmath $r$}^{\prime}|$, the integral
representation of the potential Eq.~(\ref{eq:phi}) is expanded as
follows:
\begin{equation}\label{eq:phi2}
  \phi = -\frac{GM}{r} - \frac{GM \mbox{\boldmath $r$}\cdot 
\mbox{\boldmath $R$}}{r^3}
  - \frac{G}{2}\left( 3\frac{x^i x^j}{r^5} - \frac{\delta^{ij}}{r^3}\right)
  I_{ij} + \cdots , 
\end{equation}
where 
\begin{eqnarray}
  M = \int \rho\, d^3 x, \quad
  \mbox{\boldmath $R$} = \frac{1}{M} \int \rho \mbox{\boldmath $r$}\,
  d^3 x, \quad
  I_{ij} = \int \rho x_i x_j \, d^3 x.  
\end{eqnarray}
The stationarity of the spacetime now requires that $M, 
\mbox{\boldmath $R$}, I_{ij}, \cdots$ 
are all independent of $t$. With the help of the continuity equation, 
$\partial \rho/\partial t + (\rho v^i)_{,i} = 0$, we obtain 
\begin{eqnarray}
  \int \rho v_i\, d^3 x = 0, \quad
  \int \rho v_{(i} x_{j)} \, d^3 x = 0, 
  \ \cdots .  
\end{eqnarray}
Therefore, Eq.~(\ref{eq:psi}) is expanded as
\begin{eqnarray}\label{eq:psi2}
  \psi_i = \frac{4 G  x^j}{r^3}
  \int \rho x_{[i} v_{j]}\, d^3 x  - 
  2G \left(3\frac{x^j x^k}{r^5} -
  \frac{\delta^{jk}}{r^3}\right)
  \int \rho v_k x_i x_j d^3x +\cdots.   
\end{eqnarray}
In the above equations, the round brackets denote symmetrization and
the square brackets denote skew-symmetrization. 
Finally, $\tilde{\phi}$ is expanded as 
\begin{equation}
  \tilde{\phi} = \tilde{\phi}_0 +\tilde{\phi}_1 + \tilde{\phi}_2 + \cdots, 
\end{equation}
where $\tilde{\phi}_0,\tilde{\phi}_1$ and $\tilde{\phi}_2$ represent
the monopole, dipole and quadrupole components, respectively. 
 
\subsection{Monopole and dipole contributions}

In the case of a compact rotating lens, calculation up to order 
$1/r^2$ is sufficient. Then, we find 
\begin{equation}
  \tilde{\phi}_0 + \tilde{\phi}_1 = - \frac{GM}{r} 
   - \frac{GM \mbox{\boldmath $r$}}{r^3}\cdot \left( \mbox{\boldmath $R$} 
   - \frac{\bar{\mbox{\boldmath $e$}}\times\mbox{\boldmath $L$}}{Mc}\right),  
\label{expansion}
\end{equation}
where $\mbox{\boldmath $L$}\equiv \int \rho \mbox{\boldmath $r$}\times
\mbox{\boldmath $v$} d^3x$ is the angular
momentum of the lens object.  
The dipole term can always be removed with a constant translation of the 
coordinate system: 
\begin{equation}
  \mbox{\boldmath $r$} \rightarrow  = \mbox{\boldmath $r$} 
   - \left(\mbox{\boldmath $R$} 
    - \frac{\bar{\mbox{\boldmath $e$}}\times
    \mbox{\boldmath $L$}}{Mc}\right) 
\equiv
  \mbox{\boldmath $r$} - \tilde{\mbox{\boldmath $R$}}. 
\end{equation}
Therefore, we have an important conclusion: 
A rotating lens with center of mass position $\mbox{\boldmath $R$}$ is
equivalent to a non-rotating lens with center of mass position
$\tilde{\mbox{\boldmath $R$}}$. 
The frame-dragging term due to the rotation of the lens object produces 
no additional observable effects, such as image deformation,
increase of the separation angle, or magnification of the images. 
It simply results in a constant shift of the coordinate
system. It should be noted that this result is independent of 
the assumption of a perfect fluid, since the expansion of potentials 
can always be written as Eq. ($\ref{expansion}$) with a suitable definition
of the mass and the angular momentum. 

As an alternative statement of our conclusion, we find that the 
frame-dragging effect is separable only when the ``true'' position 
of the center of mass $\mbox{\boldmath $R$}$ is independently known. 
In the following, we choose the origin of the coordinates such
that the dipole contribution vanishes: $\tilde{\phi}_1=0$.

\subsection{Quadrupole contributions}

For extended lenses, quadrupole contributions
should also be considered. 
Equations (\ref{eq:phi2}) and (\ref{eq:psi2}) give
\begin{eqnarray}
  \bar{\phi}_2 &=& - \frac{G}{2}\left( 3\frac{x^i x^j}{r^5} -
    \frac{\delta^{ij}}{r^3}\right)
  \left(I_{ij} - \frac{2}{c}\int \rho
    \bar{\mbox{\boldmath $e$}}\cdot \mbox{\boldmath $v$} x_i x_j \,d^3x
    \right) \nonumber \\
 &\equiv& - \frac{G}{2}\left( 3\frac{x^i x^j}{r^5} -
    \frac{\delta^{ij}}{r^3}\right) \tilde{I}_{ij} . 
\end{eqnarray}
Therefore, even if the ``true'' quadrupole moment of the mass
distribution vanishes,
$I_{ij}=0$, it is possible to have quadrupole contributions due to 
the effect of the dragging of inertia. 
It should be noted that to the first order in $G$, the quadrupole 
moments of the Kerr lens vanish.

\subsection{Multipole expansion of the deflection angle}

The deflection angle is given by 
\begin{equation}
  \mbox{\boldmath $\alpha$} = \frac{2}{c^2}\int_{-\infty}^{+\infty}
   \nabla \left(\tilde{\phi}_0 + \tilde{\phi}_2 + \cdots \right)\, d\ell , 
\end{equation}
where the integration is taken along the path $\mbox{\boldmath $x$} 
= \mbox{\boldmath $\xi$} + \ell \, \bar{\mbox{\boldmath $e$}}$.  
For convenience, we take the direction of 
$\bar{\mbox{\boldmath $e$}}$ as the $z$-axis, 
$\bar{\mbox{\boldmath $e$}}= (0, 0, 1)$ and 
$\mbox{\boldmath $\xi$} = (\xi^1, \xi^2, 0)$. 
Then, straightforward calculations show that $\mbox{\boldmath $\alpha$} 
= \{\alpha^i\} = (\alpha^1, \alpha^2,0)$ is
\begin{equation}
  \alpha^i = \frac{4GM}{c^2} \frac{\xi^i}{|\mbox{\boldmath $\xi$}|^2}
  -\frac{4G}{c^2}\sum_{j,k=1}^{2} \tilde{I}_{jk}
  \left(\delta^{jk} - 4 \frac{\xi^j \xi^k}{|\mbox{\boldmath $\xi$}|^2}\right)
  \frac{\xi^i}{|\mbox{\boldmath $\xi$}|^4}
  - \frac{8G}{c^2} \sum_{j=1}^{2} \tilde{I}_{ij}\frac{\xi^j}
{|\mbox{\boldmath $\xi$}|^4} . 
\end{equation}
Note that the $z$ components of the quadrupole moment $\tilde{I}_{3i}$ 
do not appear in the above equations.

\section{Summary and discussion}
We have studied the following inverse problem: If a lens is dark, 
can we infer its rotation from the observed images? 
We have found that a rotating lens cannot be distinguished from 
a non-rotating one unless the true mass density is known independently. 

In order to consider explicitly the effect due to the extended nature 
of the lens, we have used a multipole expansion analysis. 
At the dipole order, we have found that the frame-dragging term 
due to the rotation of the lens object produces 
no additional observable effects. 
It simply results in a constant shift of the coordinate
system. 
Alternatively, the frame-dragging effect is separable only when the 
``true'' position of the center of mass is independently known. 
The quadrupole order, which does not appear 
in the linearized Kerr metric, has also been studied. 
We have shown that the rotation of the lens induces a quadrupole effect 
in the gravitational lensing which results in an increase in the 
number of images. 
Even if the ``true'' quadrupole moment of the mass distribution vanishes, 
quadrupole contributions appear owing to the effect of 
dragging of inertia. 
However, we cannot recognize these additional phenomena 
as products of the rotation unless we know the true mass distribution 
through observational methods other than gravitational lensing. 

The sun is an exceptional case for which we do know the true mass
distribution. 
In this case, the light deflection caused by the rotation, quadrupole 
moment and the second order of the mass are, respectively, 
about 0.7, 0.2 and 11 $\mu$ arcsec. \cite{ES}  

In this paper, we have not considered the polarization of 
the light. In Kerr spacetime, the direction of polarization rotates 
due to the frame-dragging effect. \cite{ITT,Nouri} 
In order to make the effect of the rotation separable, we need 
additional information such as the position of the lens or 
the direction of polarization. 
Our conclusion is valid only to linear order in the gravitational constant. 
Therefore, it would be interesting to study effects at higher orders, 
though they are negligible in the astronomical situation. 

\section*{Acknowledgements}

We would like to thank H.~Ishihara, A.~Hosoya, C.~Baraldo, Y.~Kojima 
and K. Konno  for useful discussions. 
H. A. would like to thank Bernard F. Schutz for his hospitality 
at the Albert-Einstein-Institut, where a part of this work was done.
This work was supported in part by a Japanese Grant-in-Aid 
for Scientific Research from the Ministry of Education, Science 
and Culture, No. 11740130 (H. A.), and in part by the Sumitomo Foundation. 

\appendix
\section{}

Here we prove that the spatial length $\ell$ defined by
Eq.~(\ref{eq:gammaij}) is an affine parameter of the null geodesics in 
the stationary spacetime given by Eq.~(\ref{eq:llform}). 

Let us introduce the null geodesic parameterized by the affine
parameter $\lambda$. The tangent vector of the null ray is
\begin{equation}
  k^{\mu} = \frac{dx^{\mu}}{d\lambda} . 
\end{equation}
The geodesic equation for $k_{\mu}$ is 
\begin{equation}
  \frac{dk_{\mu}}{d\lambda} =
  \frac{1}{2} g_{\alpha\beta, \mu} k^{\alpha} k^{\beta} . 
\end{equation}
Since the spacetime is stationary, the metric does not depend on the
time coordinate $x^0$. Therefore, $k_0$ is constant along the
geodesic: $dk_0/d\lambda = 0$. 
The null  condition $k^{\mu} k_{\mu} = 0$ reads 
\begin{equation}
  \left(k_0\right)^2 =
  \gamma_{ij} \frac{dx^i}{d\lambda}\frac{dx^j}{d\lambda}=
  \left(\frac{d\ell}{d\lambda}\right)^2. 
\end{equation}
Hence, 
\begin{equation}
  \frac{d}{d\lambda}\left(\frac{d\ell}{d\lambda}\right) = 0, 
\end{equation}
which implies that $\ell$ is another affine parameter.


\begin{thebibliography}{99}
\bibitem{SEF} For example, P. Schneider, J. Ehlers and E. E. Falco, {\it
    Gravitational Lenses\/} (Springer-Verlag, Berlin, 1992). 
\bibitem{ES} R. Epstein and I. I. Shapiro, Phys. Rev. {\bf D22} 
(1980), 2947. 
\bibitem{Ibanez} J. Ibanez, Astron. Astrophys. {\bf 124} (1983), 175. 
\bibitem{Sari} S. O. Sari, Astrophys. J. {\bf 462} (1996), 110. 
\bibitem{KK} K. Konno and Y. Kojima, Prog. Theor. Phys. 
{\bf 101} (1999), 885. 
\bibitem{BHN} C. Baraldo, A. Hosoya and T. T. Nakamura, 
  Phys. Rev. {\bf D59} (1999), 083001. 
\bibitem{LL} L. D. Landau and E. M. Lifshitz, {\it The Classical Theory 
    of Fields} (Oxford: Pergamon 1962). 
\bibitem{Perlick} V. Perlick, Class. Quant. Grav. {\bf 7} (1990), 1319. 
\bibitem{YO} H. Yoshida and M. Omote, Prog. Theor. Phys. {\bf 79} 
(1988), 1095. 
\bibitem{Durrer} R. Durrer, Phys. Rev. {\bf D72} (1994), 3301. 
\bibitem{ITT} H. Ishihara, M. Takahashi and A. Tomimatsu, 
  Phys. Rev. {\bf D38} (1988), 472.
\bibitem{Nouri} M. Nouri-Zonoz, Phys. Rev. {\bf D60} (1999), 024013. 
\end{thebibliography}
\end{document}